\documentclass[%
 onecolumn,
superscriptaddress,
 amsmath,amssymb,
 aps,
longbibliography
]{revtex4-1}

\usepackage{graphicx}
\usepackage{dcolumn}
\usepackage{bm}
\usepackage[mathlines]{lineno}

\usepackage{siunitx}

\def \vavg{\langle v_{CM} \rangle}

\usepackage{color}

\begin{document}

\preprint{APS/123-QED}

\title{Kinematics of a simple reciprocal model swimmer at intermediate Reynolds numbers}

\author{Thomas Dombrowski}
 \affiliation{Department of Physics, University of North Carolina at Chapel Hill, Chapel Hill, NC, USA}
\author{Daphne Klotsa}%
 \email{dklotsa@email.unc.edu}
\affiliation{%
Department of Applied Physical Sciences, University of North Carolina at Chapel Hill, Chapel Hill, NC, USA\\
}%


\date{\today}

\begin{abstract}
We computationally study the kinematics of a simple model reciprocal swimmer (asymmetric dumbbell) as a function of the Reynolds number (Re) and investigate how the onset and gradual increase of inertia impacts the swimming behavior: a reversal in the swim direction, flow directions, and the swim stroke. We divide the swim stroke into the expansion and compression of the two spheres and relate them to power and recovery strokes.  We find that the switch in swim direction also corresponds to a switch in power and recovery strokes. We obtain expressions for the mean swimming velocity by collapsing the net displacement during expansion and compression under power law relationships with respect to Re, the swimmer's amplitude, and the distance between the two spheres. Analyzing the fluid flows, we see the averaged flow field during expansion always resembles a pusher and compression always a puller, but when averaged over the whole cycle, the flow that dominates is the one that occurs during the power stroke. We also relate the power and recovery strokes to the swimming efficiency during times of expansion and compression, and we find that the power stroke is, surprisingly, not always more efficient than the recovery stroke. Our results may have important implications for biology and ultimately the design of artificial swimmers.

\begin{description}
\item[PACS numbers]
May be entered using the \verb+\pacs{#1}+ command.
\end{description}
\end{abstract}

\pacs{Valid PACS appear here}
\maketitle

\section{\label{sec:Intro}Introduction}

Biological and artificial swimmers exist across a broad range of length scales, spanning from micron-sized bacteria and self-propelled nanoparticles to large aquatic organisms and marine robots  on the order of meters. Swimming can be categorized by the Reynolds number (Re) which relates viscous and inertial forces. Microscopic swimmers at low Re, where viscosity dominates, swim differently than high-Re swimmers, where inertia dominates. Indeed, in nature one can see bacteria swim with a corkscrew chiral flagellum at low Re, while larger fish undulate their bodies pushing fluid backwards to move forwards at high Re. Between the two extremes resides the intermediate Reynolds regime, where both viscosity and inertia play a role. Mesoscopic organisms i.e. those that operate at intermediate Re are diverse both in size, $\approx$ 0.5mm - 50cm, and in swimming mechanisms, including for example jet propulsion of squid and jellyfish~\cite{Bartol2009,Herschlag2011}, rowing of copepod antennae~\cite{Strickler1975,Blake1986}, aquatic flapping flight of pteropods~\cite{Borrell2005,Mohaghar2019}, anguilliform (eel-like) locomotion~\cite{Kern2006,fuiman1988ontogeny,Sznitman2010,McHenry2003,Bhalla13FD}, and ciliate beating~\cite{gemmell2015tale,jiang2011does}. 
Understanding motility in fluids is important both for answering fundamental biological questions, such as how do organisms swim, feed, communicate, etc. but also for the design of artificial swimmers and flyers, such as marine robots and drones. 

To gain insight into generic features and underlying physical mechanisms, simple theoretical models have been developed such as the scallop and Purcell's three-link-swimmer~\cite{Purcell1977}, the squirmer model~\cite{lighthill1952squirming,blake1971squirmer,Lauga2009, Pedley16}, asymmetric and symmetric dumbbell swimmers~\cite{alexander2008dumb,lauga2008no,putz2010low}, the three-sphere swimmer~\cite{Najafi2004}, and the push-me-pull-you swimmer~\cite{Avron2005}. Most of the models have focused on microscopic scales where inertia is negligible because a) there are a lot of interesting biological questions and applications at microscopic scales, such as intracellular dynamics and processes in the cytoplasm, cell motility, bacteria, and algae~\cite{Goldstein2016batchelor}, as well as artificial swimmers, such as self-propelled colloids and nanoparticles aspiring for example to aid in drug delivery~\cite{mallory2018active,din2016}. And b) because Stokesian swimmers must break time reversibility, which makes their design theoretically challenging. While the Stokes regime is indeed very interesting, it is as important to understand what happens as we move away from the strict $\text{Re} = 0$ Stokes regime, when and how inertia kicks in, and its consequences for different kinds of swimmers (e.g. different geometries and motility mechanisms). Models that include finite inertia are the inertial squirmer~\cite{lauga-continuous,arezoodaki_2012inertial,khair2014expansions,chisholm16,Li2016,chisholm2018partial}, the flapping plate~\cite{zhang2010locomotion,spagnolie2010surprising}, and the asymmetric and symmetric dumbbell swimmers~\cite{Klotsa2015,Felderhof2016,Collis2017, dombrowski2019transition, Parthasarathy2019}. 

It is worth noting that a lot of biology takes place near the boundary between the Stokes and intermediate Reynolds regimes, yet where the boundary is precisely is generally unknown. Where the boundary is matters because organisms have to change their swimming mode, feeding strategy, etc. depending on the regime in which they live. Switching regimes is not unusual, in fact, a plethora of organisms born into the Stokes regime move out of it as they grow in size.We would expect that they also change the way they move as a result of this change in regime.  
For example, the mollusk \textit{C.~antartica} switches from using cilia to flapping as it grows \cite{childress2004transition}, the brine shrimp transitions from rowing to gliding with metachronally-beating legs~\cite{williams1994model}, and the nymphal mayfly transitions from rowing to flapping with its gill plates~\cite{sensenig2009rowing}. 
From an applications point of view, understanding the physics near the boundary can help us design artificial swimmers or (microfluidic) processes that utilize the relative ratio of inertial and viscous forces, switching between regimes, and thus switching between desired properties. For a longer discussion on biology and applications at intermediate Re, see~\cite{Klotsa2019Perspective}.

In this paper, we studied the kinematics of a simple reciprocal model swimmer as a function of the Reynolds number. The same asymmetric dumbbell model (termed the spherobot) was determined to switch swim direction depending on the Reynolds number because of the corresponding induced steady streaming flows~\cite{dombrowski2019transition}. The spherobot switched from a small-sphere-leading regime to a large-sphere-leading regime at $\text{Re}_c \approx 20$. Here, we studied the motion of the spherobot swimmer in more detail by splitting its oscillation into the expansion and compression of the two spheres and collapsed their corresponding net displacements under piece-wise power law relationships with respect to $\text{Re}$, inverse Strouhal number $\epsilon$, and equilibrium distance between spheres $d_0$.
We also related the expansion and compression to power and recovery strokes. We found that the switch in swim direction as Re increased corresponded to a switch in the power and recovery strokes. In the small-sphere leading regime ($\text{Re}<\text{Re}_c$), the power stroke occurred during compression and the recovery stroke during expansion, while the reverse occurred in the large-sphere-leading regime ($\text{Re} > \text{Re}_c$). We noticed how as Re increased and inertial forces became more dominant, our swimmer transitioned from a jerky, back-and-forth motion with a large backward displacement during the recovery stroke in the small-sphere-leading regime to a continuous movement forward in the direction of swimming all in the same direction, with no backward displacement during the recovery stroke in the large-sphere-leading regime. By studying the fluid flows, we saw that the averaged flow field during expansion was always pusher-like and during compression puller-like, which is to be expected, but when averaged over the whole cycle one of the two flow fields dominated. We determined the most dominant flows consistently occurred during the power stroke in each regime. We also related the power and recovery strokes to the spherobot's efficiency during times of expansion and compression, and we found that the power stroke was, surprisingly, not always more efficient than the recovery stroke. The subtle differences in Re that can lead to switching regimes and swim strokes may have important implications for biology and ultimately the design of artificial swimmers. 

The structure of the paper is as follows. In section II, we briefly describe the model, computational method, and simulation details. In section III we present results for the kinematics of the spherobot, section IV for averaged fluid flows, efficiencies, and the evolution of fluid flow. We end with discussion and conclusions in section V.

\section{\label{sec:ModelMethods}Model, methods, and background}

\begin{figure}
\includegraphics[width=0.8\columnwidth]{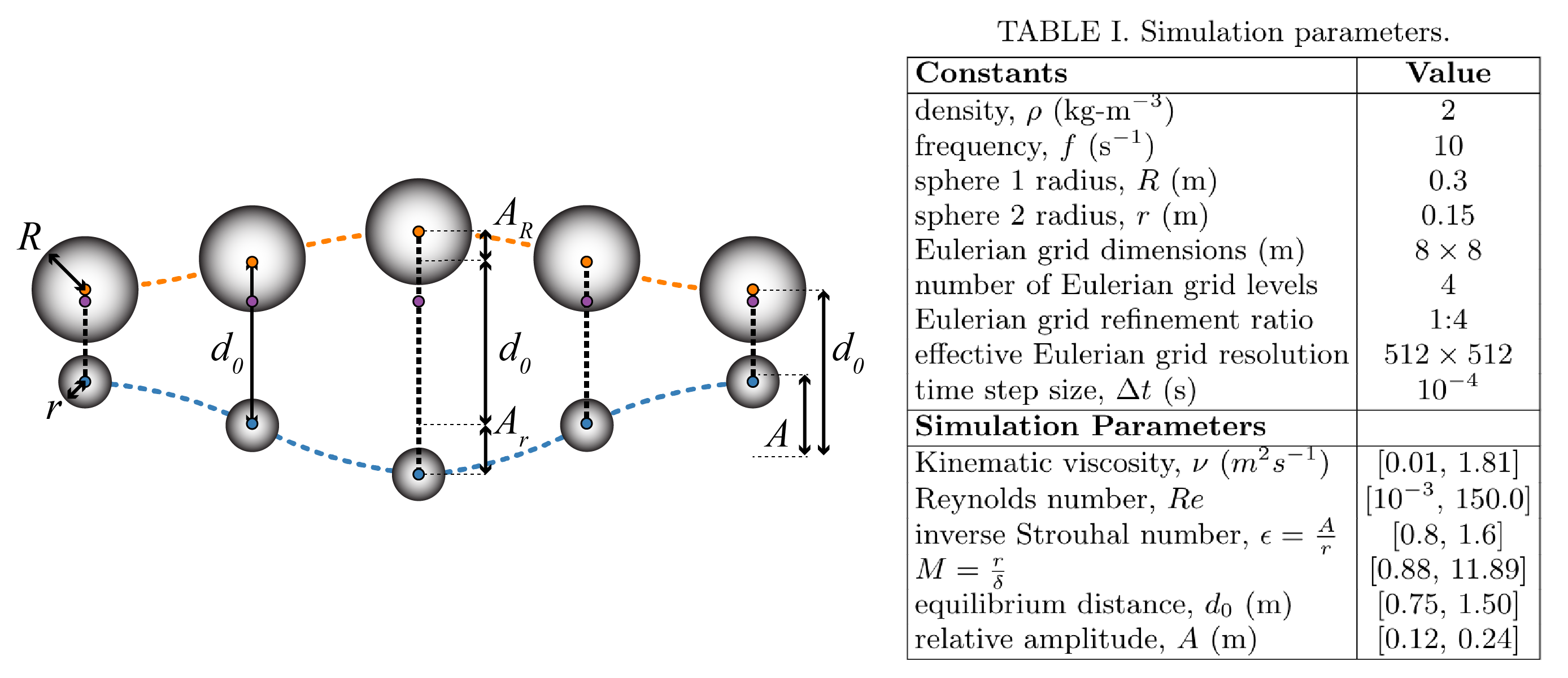}
\centering
\caption{Left: Reciprocal oscillation of the spherobot swimmer over one cycle. The large sphere (orange) with radius $R$ always oscillates in the opposite direction of the small sphere (blue) with radius $r$. The distance between the spheres $d(t) = d_0 + Asin(2\pi t)$ is prescribed to be of a simple harmonic oscillator with frequency $f$, where $d_0$ is the equilibrium distance between the spheres, and $A = A_r + A_R$ is the amplitude of the spherobot. The amplitude of the small and large spheres are $A_r$ and $A_R$ respectively. When absent of fluid, the spherobot's center of mass (CM), shown in purple, does not move throughout the oscillation. 
}
\label{diagram}
\end{figure}

The spherobot is a geometrically simple, reciprocal model swimmer composed of two unequally sized spheres of radii $R$ and $r$, such that $R > r$, see also~\cite{dombrowski2019transition,SM}). The spheres oscillate in antiphase with respect to each other, and they are coupled to one another by prescribing the distance between their centers, $d(t)=d_0+A \sin(2\pi ft)$, with an actuated spring, where $d_0$ is the equilibrium distance between the centers, $A$ is the amplitude of the spherobot $A = 0.5(d_{max} - d_{min})$, and $f$ is the frequency of oscillation, all of which are depicted in Fig.~\ref{diagram}. As a result, an equal and opposite (spring) force was applied to each sphere and moved them the prescribed distance apart, $F_{R} = - F_{r}$. Since the forces were equal in magnitude and the spheres were of the same density, their amplitudes were different: $A_R < A_r$ and $A = A_R + A_r$. Subscripts $R$ and $r$ indicate quantities specific to the large and small sphere, respectively. Both spheres were neutrally buoyant with the surrounding fluid, i.e. they had equal densities $\rho_p = \rho_f = \rho$. 

The spherobot was immersed in a viscous, incompressible fluid that occupied a finite cell with no-slip walls. The fully-coupled fluid-structure interaction system was resolved using 
the Constrained Immersed Boundary (CIB) method~\cite{Kallemov16,Usabiaga17}. The CIB scheme was implemented in IBAMR, which is an immersed boundary numerical method with adaptive mesh refinement~\cite{griffith2007adaptive, IBAMR-web-page}. An adaptive mesh is implemented to improve the efficiency of the simulation. The coarsest level is broken up into N=8 cells along one dimension. We use a grid refinement ratio of 1:4 where the next highest refinements are N=32, 128, and 512. There are 4 refinement levels, and the spherobot's mesh is evaluated at the highest grid refinement of N=512. The simulation box is 6$\times$ the length of the swimmer and it is 53$\times$ the radius of the small sphere to prevent interactions with the wall. 

In previous work, we investigated the spherobot, which was shown to switch swim direction depending on a critical Re~\cite{dombrowski2019transition}. The swim direction was related to the reversal of steady streaming flows (SS) around the small sphere~\cite{riley66}. We note three important findings that are relevant in this paper too. 1) The steady streaming reversal of the time-averaged flow fields over a cycle was qualitatively similar to puller and pusher flow fields defined in Stokes flows, although these swimmers were at finite inertia and no assumptions were made on the fluid flows. 2) We showed that steady streaming flows can be used as a propulsion mechanism, which is especially interesting in the low-intermediate Re $<20$, where inertia is weak. And 3) Although the SS flows reversal is gradual as a function of Re, the result in a system like ours is a drastic change in behavior, namely change the direction of swimming. 

Before we further discuss the spherobot, let us first consider the simpler problem of a single oscillating sphere with angular frequency $\omega$, amplitude $A$, and radius $r$. In general, there are three relevant Reynolds numbers~\cite{lauga-continuous}: the particle one $\text{Re}_p=\rho_p r^2 \omega / \rho_f \nu$, the frequency one $\text{Re}_f=rA\omega/\nu$ which scales with the non-linear advective term in the Navier-Stokes equations~\cite{vandenberghe2006unidirectional,vandenberghe2004symmetry,alben2005coherent,lauga-continuous}, and $M^2=r^2\omega/\nu$, which scales with the unsteady term in the Navier-Stokes equations~\cite{riley66}. The oscillatory motion introduces a relevant length scale, the oscillatory boundary layer thickness, $\delta=\sqrt{\nu/\omega}$~\cite{boundary-layer-book}. When there are steady streaming flows there is an additional dimensionless ratio, the streaming Reynolds number $Re_s = A^2\omega/\nu$, which quantifies the steady streaming flows around a single oscillating sphere beyond the oscillatory boundary layer~\cite{riley66,riley2001,chang1994unsteady}. Thus, even for the simpler case of one oscillating sphere there are at least four relevant Re. 

Additional complexity enters the system when we include a second sphere of different size oscillating antiphase, resulting in net motion, i.e. swimming. First, there are the extra length scales of the second sphere and of the whole swimmer, as well as the amplitude of the second sphere (apart from the ones stated above i.e. amplitude and radius of first sphere, distance between spheres, oscillatory boundary layer thickness). Second, there is a swimming Reynolds $\text{Re}_{swim}=Ul/\nu$, where $U$ is the swim velocity, and $l$ the swimmer's length scale. It is worth noting that for experimentally relevant systems, many of these length scales (that enter the different dimensionless ratios) are of the same order of magnitude giving values for the ratios close to 1. As a result, the problem becomes more challenging and often analytically intractable~\cite{coenen}. 

Here, we use $\text{Re} = A_r r\omega/\nu$ as the reference Reynolds number (for simplicity the notation will be just $\text{Re}$) because that was the Re we found to determine the spherobot switch in swim direction from a small-sphere-leading regime (SSL) to a large-sphere-leading regime (LSL)~\cite{dombrowski2019transition}. Note that a similar Re has been shown to dominate other intermediate-Re phenomena: the scaling of the stagnation point indicating the reversal of outer and inner steady streaming~\cite{tatsuno1981Unharmonic,kotas2007visualization} and the gap between two granular spheres oscillating in phase~\cite{klotsa2007,klotsa2009}. 



We investigated the spherobot's movement in Stokes flow and in the range of $0.5 \leq Re \leq 150$. We performed a parameter sweep varying the fluid's kinematic viscosity $\nu$, the spherobot's amplitude $A$, the equilibrium distance between spheres $d_0$, while keeping the spheres' radii $r$ and $R$, frequency $f$, and sphere and fluid density $\rho$ constant ($r$ = 0.15m, $R$ = 0.3m, $f=10$Hz, and $\rho = 2$kg/$\text{m}^3$).
All parameters are shown in Table 1. 
The simulations were run long enough for the spherobot to reach a steady state, defined as less than a one percent change in the average velocity over consecutive oscillations. Data was acquired after steady state was reached. In most of the paper we focused on two characteristic systems, one in the small-sphere-leading regime at $\text{Re}=2.5$ and one in the large-sphere-leading regime at $\text{Re}=70.0$. For both systems, $d_0$ = 0.975m  and $A$ = 0.18m. We used the software VisIt~\cite{HPV:VisIt} for fluid flow analysis. Other analysis was done with in-house Python code.
In the rest of the paper, we assume the spherobot is placed vertically (y-direction) with the large sphere on top and the small sphere at the bottom (as shown in Fig.~\ref{diagram}). For all figures showing a characteristic cycle of oscillation, the data is shifted in the time axis in the following way. The first half of the cycle is a region of expansion, followed by a region of compression in the second half. We  define $\tau = ft$ as our dimensionless time unit, essentially the fraction of time elapsed in the cycle. At $\tau=0.00, 1.00$ the spheres are at minimum distance $d_0 - A$, at $\tau=0.50$ they are at maximum distance $d_0 + A$, and at $\tau=0.25, 0.75$ they are at their equilibrium distance apart $d_0$. 

\section{\label{sec:COM}Results}
\subsection{Kinematics}

\begin{figure}
\includegraphics[width=0.75\columnwidth]{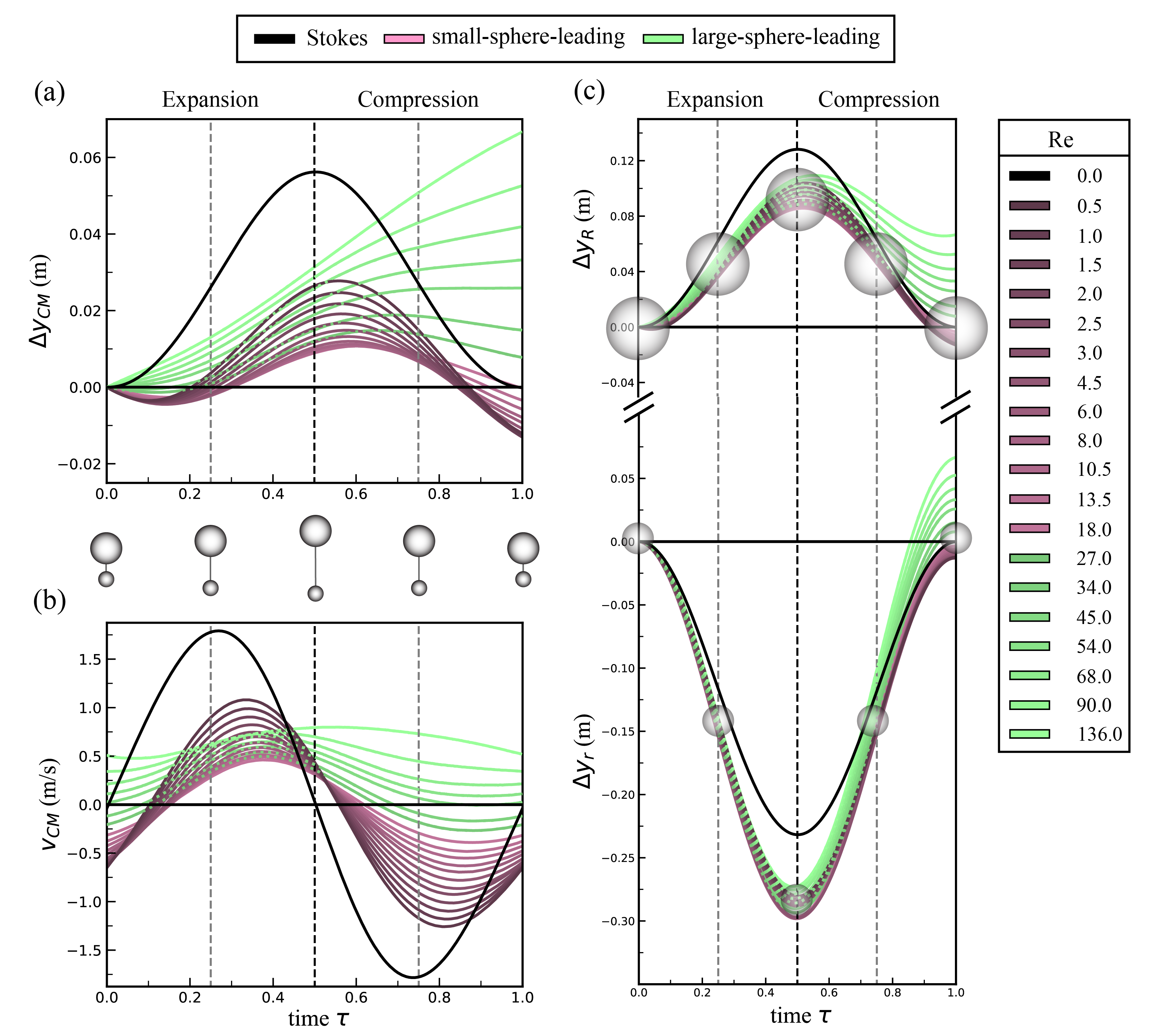}
\centering
\caption{Kinematic quantities plotted as functions of time over one cycle of oscillation after steady state had been reached. (a) Displacement of spherobot, represented by the net displacement of its center of mass $\Delta y_{CM}$, (b) velocity of spherobot represented by $v_{CM}$, and (c) displacement of individual spheres of radius $R$ and $r$. Net swimming direction is indicated by the colors of the curves (black = Stokes flow, pink = small-sphere-leading, and green = large-sphere-leading. The Reynolds number is represented by the shading of the curves, see legend.
}
\label{COMFig}
\end{figure}

We first studied how the periodic oscillation of the two spheres, that composed the spherobot, resulted in net displacement of their combined center of mass (CM) over one cycle, $y_{CM} = (y_r m_r + y_R m_R) / (m_r + m_R)$, where $y$ indicates position along the swimmer's axis and $m$ the mass of each sphere indicated by the subscript, for a range of Re (Fig.~\ref{COMFig}). Note that because of the unequal masses of the spheres, the CM is actually on the large sphere (see Fig.~\ref{diagram}) and as such closely follows the trajectory of the large sphere. We used the CM to indicate the displacement and velocity of the spherobot as a whole. Moreover, displacement was measured in relation to the position of the CM at the start of the cycle at $y_{CM} =0$. The full parameter range of data shown in Table 1 is found in the SI. We present our findings where the spherobot's amplitude, $A = 0.18m$, the equilibrium distance between spheres, $d_0=0.975m$, and the individual sphere radii, $R = 0.3m$ and $r = 0.15m$, were held constant such that Re was only $\propto 1/\nu$. In other words, Re was increased gradually via the kinematic viscosity $\nu$.   

For $\text{Re}=0$ (Stokes flow), the spherobot's reciprocal motion resulted in no net displacement over a cycle, as expected from the scallop theorem~\cite{Purcell1977,Lauga.Bartolo.2008}. It moved in the direction of the large sphere during expansion, reaching maximum displacement half way through the cycle and then moved in the direction of the small sphere during compression, ultimately returning exactly where it began, see Fig.~\ref{COMFig}(a) black curve. 

\subsubsection{Small-sphere-leading (SSL)}
As we transitioned from Stokes flow to intermediate Re, the spherobot's trajectory changed, see Fig.~\ref{COMFig}(a) pink curves. At the start of its cycle, the spherobot moves forward (small sphere on the front), then slightly in the opposite direction during expansion and the initial part of compression; it moves with the small sphere on the front for the rest of the compression, with net displacement in that same direction at the end of the cycle. Note that the maximum displacement during the cycle is in the opposite direction to that of net swimming. This backward maximum displacement occurred at the half-period mark for Stokes flows and was shifted to a later time $\tau \approx 0.55 - 0.65$ in the small-sphere-leading regime. As Re increased both the maximum backward displacement near the half-period mark and the net displacement at the end of the cycle got smaller, see curves from Re=0.5 to Re=13.5, at $\tau\approx 0.5$ and $\tau=1$, respectively. For Re=18.0, the net displacement after one cycle is $\approx 0$. The spherobot will switch direction and transition from the small-sphere-leading to the large-sphere-leading regime.  

\subsubsection{Large-sphere-leading (LSL)}
 In the large-sphere-leading regime we see two behaviors, see Fig.~\ref{COMFig}(a) green curves. First, for Re=27 and Re=34, at the start of its cycle, the spherobot moves backward slightly (small sphere on the front), and then moves forward (large sphere on the front) during expansion. It then continues to move forward during compression, only to slightly move back again at the end of the compression, with net displacement towards the large sphere. Already, it is clear that in the large-sphere-leading regime, the spherobot is hardly ever found to be with displacement in the opposite direction to its swimming, converse to the small-sphere-leading regime. Then, as Re increases further ($\text{Re} > 45$), the backward motion is suppressed more until the spherobot moves in the direction of swimming at all times. The two behaviors are more evident from the velocity plots, see Fig.~\ref{COMFig}(b) green curves, where for Re=27 and Re=34 the velocity at the start and the end of the cycle is negative (towards the small sphere), while for all other higher Re, the velocity is always positive (towards the large sphere). 
 
 \begin{figure}
\includegraphics[width=0.55\columnwidth]{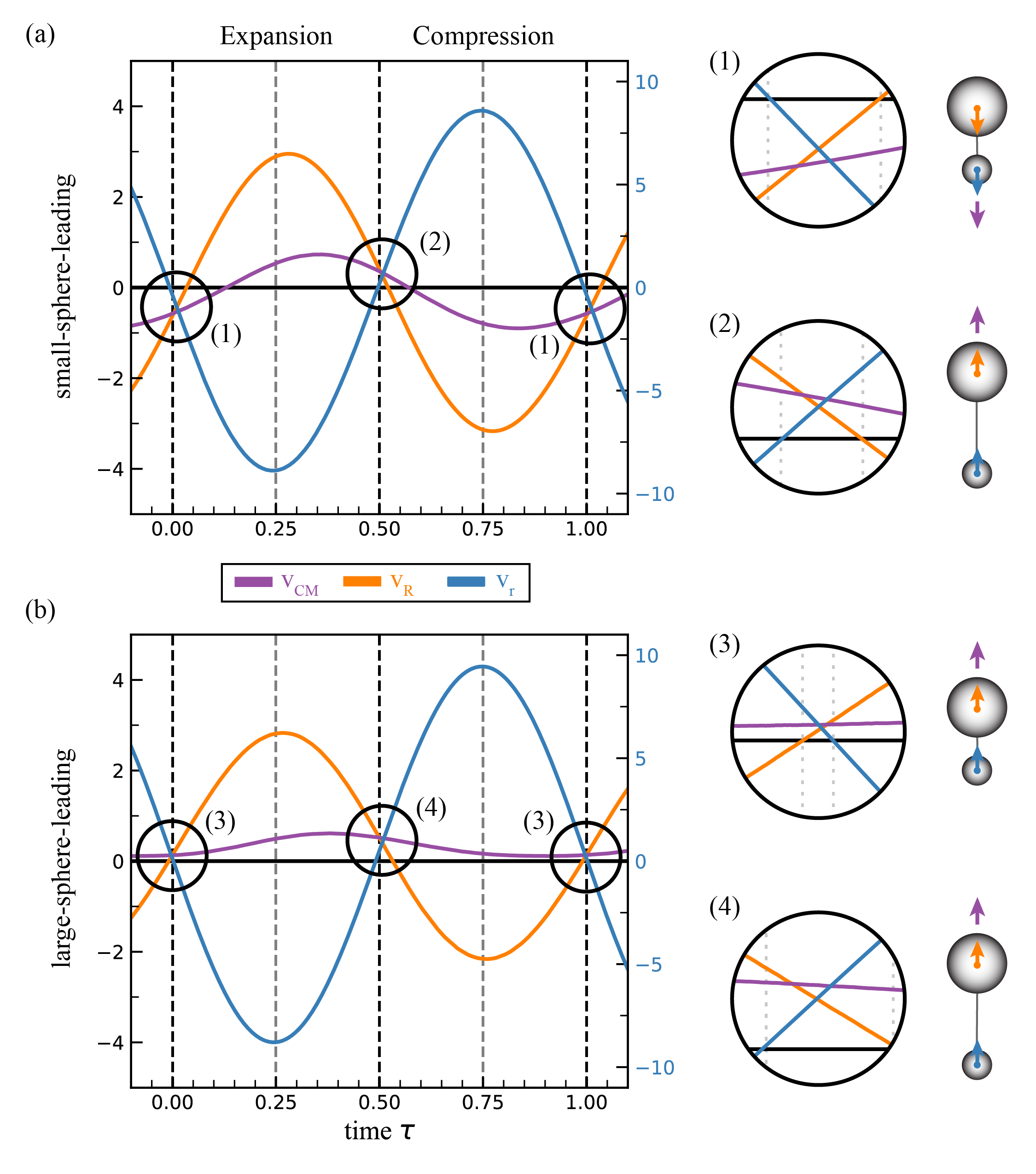}
\centering
\caption{We define slip to be the region of movement where both spheres move in the same direction. As a result, the entire spherobot moves in the same direction as its spheres. Here, we show the velocities of the spherobot $v_{CM}$ (purple), its large sphere $v_R$ (orange), and its small sphere $v_r$ (blue) measured in (m/s) when it is (a) small-sphere-leading and (b) large-sphere-leading. We identify the regions of slip observed during each spherobot's oscillation with black circles, and the region is magnified to the right. For the (a) small-sphere-leading spherobot, slip regions (1) and (2) are shown. (1) displays a region where slip is small-sphere-leading. It occurs at the end of compression and at the start of expansion. (2) shows a region where slip is large-sphere-leading. It occurs at the end of expansion and the start of compression. For the (b) large-sphere-leading spherobot, slip regions (3) and (4) are identified. (3) and (4) show regions where the spherobot slips large-sphere-leading. (3) occurs at the end of compression and the start of expansion. (4) occurs at the end of expansion and the start of compression.
}
\label{VelComparison}
\end{figure}

\subsubsection{Separate spheres}

To understand how each sphere contributes to the overall motion, we also looked at the kinematics of the spheres separately, see Fig.~\ref{COMFig}(c). During expansion (compression), for both regimes the large sphere's net displacement $\Delta y_{R}$ is always LSL (SSL), and the small sphere's $\Delta y_{r}$ is always SSL (LSL). During compression, the large and small spheres do the opposite. The distinction in the trajectories of the two regimes seems to appear during compression.

In Stokes flow, the trajectory of each sphere is symmetric with respect to time over a cycle, and the two spheres are always antiphase. As we increase Re, the individual spheres are affected by the onset of inertia differently resulting in a phase difference between them. We present data for two characteristic systems (described in the Methods section), one in each regime. In Fig.~\ref{VelComparison}, we compared the velocities of the large sphere (orange), small sphere (blue), CM (purple), and identified regions of ``slip'' to be when both spheres moved in the same direction. When both spheres' velocities are negative (toward the small sphere) we refer to SSL-slip and when they are both positive (toward the large sphere), we refer to LSL-slip. 
In the small-sphere-leading regime, at the end of expansion and the start of compression we found slip in the direction opposite to swimming (LSL-slip), while at the end of compression and the start of expansion we found a larger slip in the direction of swimming (SSL-slip), see Fig.~\ref{VelComparison}(a).
In the large-sphere-leading regime, at the end of expansion and the start of compression the slip was still LSL but now in the direction of swimming, while at the end of compression and the start of expansion we found that the direction of slip depended on the Re. As Re increased, the slip switched to LSL. In other words, the increase in inertia only affected the slip direction after compression. So, we identified for $\text{Re} > 0.0$ two contributions to the motion of the spherobot, the oscillatory and the slip (steady).

We also varied $A$ and $d_0$ in addition to $\nu$, shown in Fig.~\ref{COMFig}, the full parameter range shown in Table 1, and additional plots are included in the SI (section I). The magnitude of the net displacement at the end of the cycle increased when $A$ and $\text{Re}$ also increased. Vice versa, the net displacement decreased when $d_0$ increased (Fig. S1,S2).

\begin{figure}
\includegraphics[width=0.75\columnwidth]{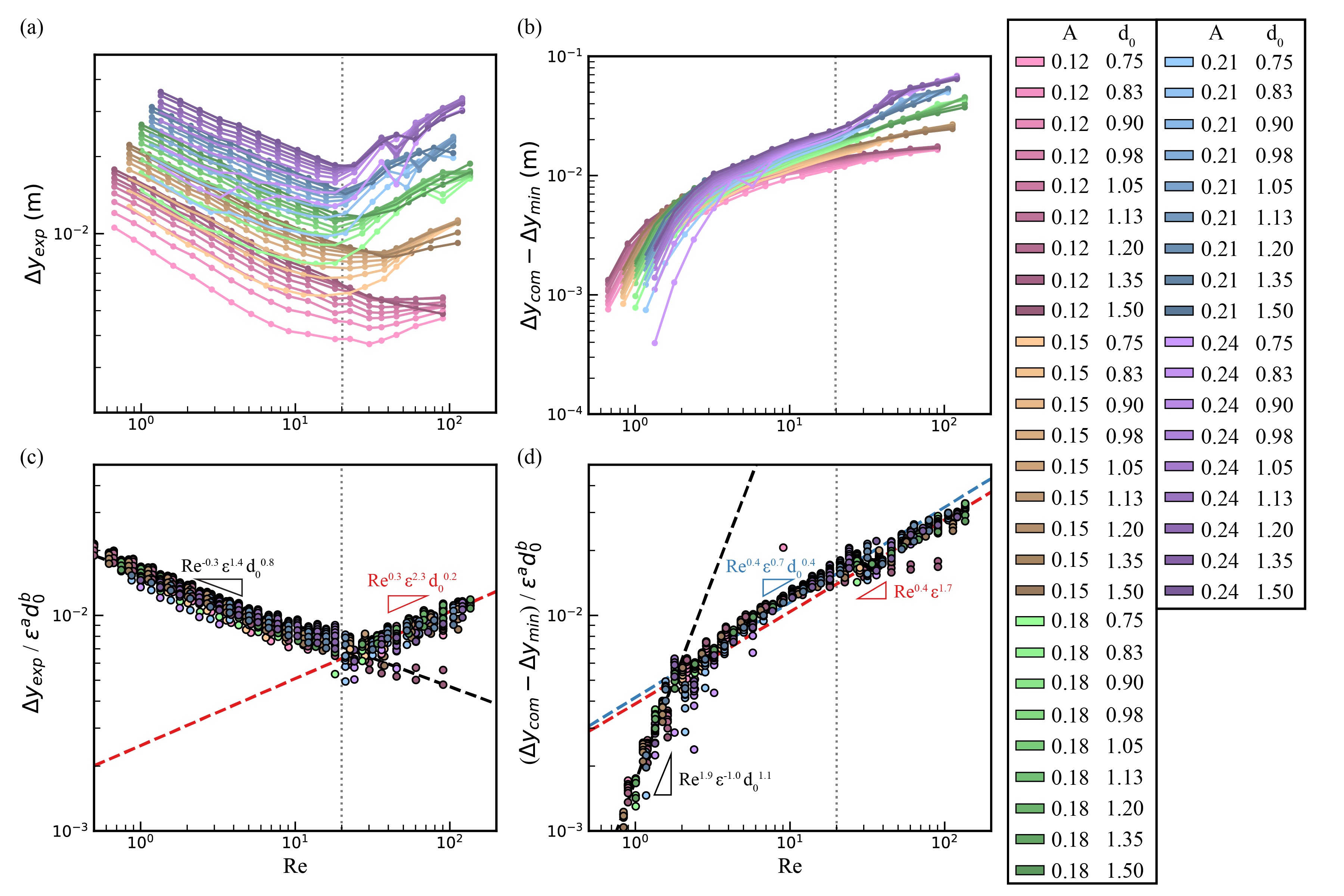}
\centering
\caption{
(a) Net displacement of the spherobot during expansion $\Delta y_{exp}$ as a function of $\text{Re}$ on a log-log scale for all $A$ and $d_0$ simulated, see legend. All curves show a constant negative slope followed by a positive slope, the turning point and the positive slope change for different amplitudes $A$ (color). (b) Net displacement of the spherobot during compression $\Delta y_{com}$ as a function of $\text{Re}$ on a log-log scale. There are three distinct positive slope trends with $\text{Re}$. We fit the expansion and compression displacements with respect to $\text{Re}$, $\epsilon$, and $d_0$ using a multiple variable linear regression. (c) $\Delta y_{exp}$  vs. $\text{Re}$ on a log-log scale collapsed into a negative slope (black dashed) and a positive slope region (red dashed). (d) $\Delta y_{com}$ vs. $\text{Re}$ on a log-log scale collapsed into three distinct positive slope regions. The corresponding relationships with $\text{Re}$, $\epsilon$, and $d_0$ are also shown in (c) and (d).
}
\label{Fit}
\end{figure}

To help identify trends in the data, we decomposed $\Delta y_{CM}$ into the net displacements during expansion $\Delta y_{exp}$ and during compression $\Delta y_{com}$. Fig.~\ref{Fit}(a,b) shows an example of this decomposition for all amplitudes and equilibrium distances studied. 
Let us consider expansion first. 
When the $\Delta y_{exp}$ data is plotted on a log-log scale, see Fig.~\ref{Fit}(a), we see a constant negative slope followed by a constant positive slope, indicative of two regions, each defined by a power law in \text{Re}. 
If we now consider compression and look at the data for $\Delta y_{com}$ on a log-log scale, see Fig.~\ref{Fit}(b), we observe three distinct trends with respect to \text{Re} all with positive slopes, also determined to be power laws with different exponents. 
We considered three variables $\text{Re}$, $\epsilon=r/A$, and $d_0$ and assume they are independent of one another.
We partitioned the data into two expansion regions and three compression regions with \text{Re}, and performed a multiple variable linear regression on each. It is important to note that each region has a different dependence on $\text{Re}$, $\epsilon$, and $d_0$. For the expansion, the data was split where there was a minimum in $\Delta y_{exp}$, see Fig.~\ref{Fit}(a).
For compression, the data was split where the slope changed at $\text{Re} \approx 2.0$ and then again when $\Delta y_{exp}$ was at a minimum (the same criterion as the expansion data), see Fig.~\ref{Fit}(b).
The resulting collapse is shown on a log-log scale in Fig.~\ref{Fit}(c,d). 
Equations~(\ref{ExpFit}) and (\ref{ComFit}) show the fits for $\Delta y_{exp}$ and $\Delta y_{com}$, respectively, and their power law relationships with $\text{Re}$, $\epsilon$, and $d_0$. It is worth noting that while there is currently no analytical theory for finite amplitudes, the expressions we obtained from the collapse can be used to give a prediction for the velocity of the spherobot, $\vavg = f(\Delta y_{exp} + \Delta y_{com})$, where $f$ is the frequency of its oscillation.  

\begin{equation}\label{ExpFit}
\Delta y_{exp} \text{ =}
	\begin{cases}
      \text{10}^{-1.8}\text{Re}^{-0.3}\epsilon^{1.4}d_0^{0.8} & \text{Fig.~\ref{Fit}(c) Black} \\
      \text{10}^{-2.6}\text{Re}^{0.3}\epsilon^{2.3}d_0^{0.2} & \text{Fig.~\ref{Fit}(c) Red} \\
   \end{cases}
\end{equation}

\begin{equation}\label{ComFit}
\Delta y_{com} \text{ =}
	\begin{cases}
      \text{10}^{-2.8}\text{Re}^{1.9}\epsilon^{-1.0}d_0^{1.1} & \text{Fig.~\ref{Fit}(d) Black} \\
      \text{10}^{-2.4}\text{Re}^{0.4}\epsilon^{0.7}d_0^{0.4} & \text{Fig.~\ref{Fit}(d) Blue} \\
      \text{10}^{-2.4}\text{Re}^{0.4}\epsilon^{1.7} & \text{Fig.~\ref{Fit}(d) Red}
   \end{cases}
\end{equation}

\begin{figure}
\includegraphics[width=0.5\columnwidth]{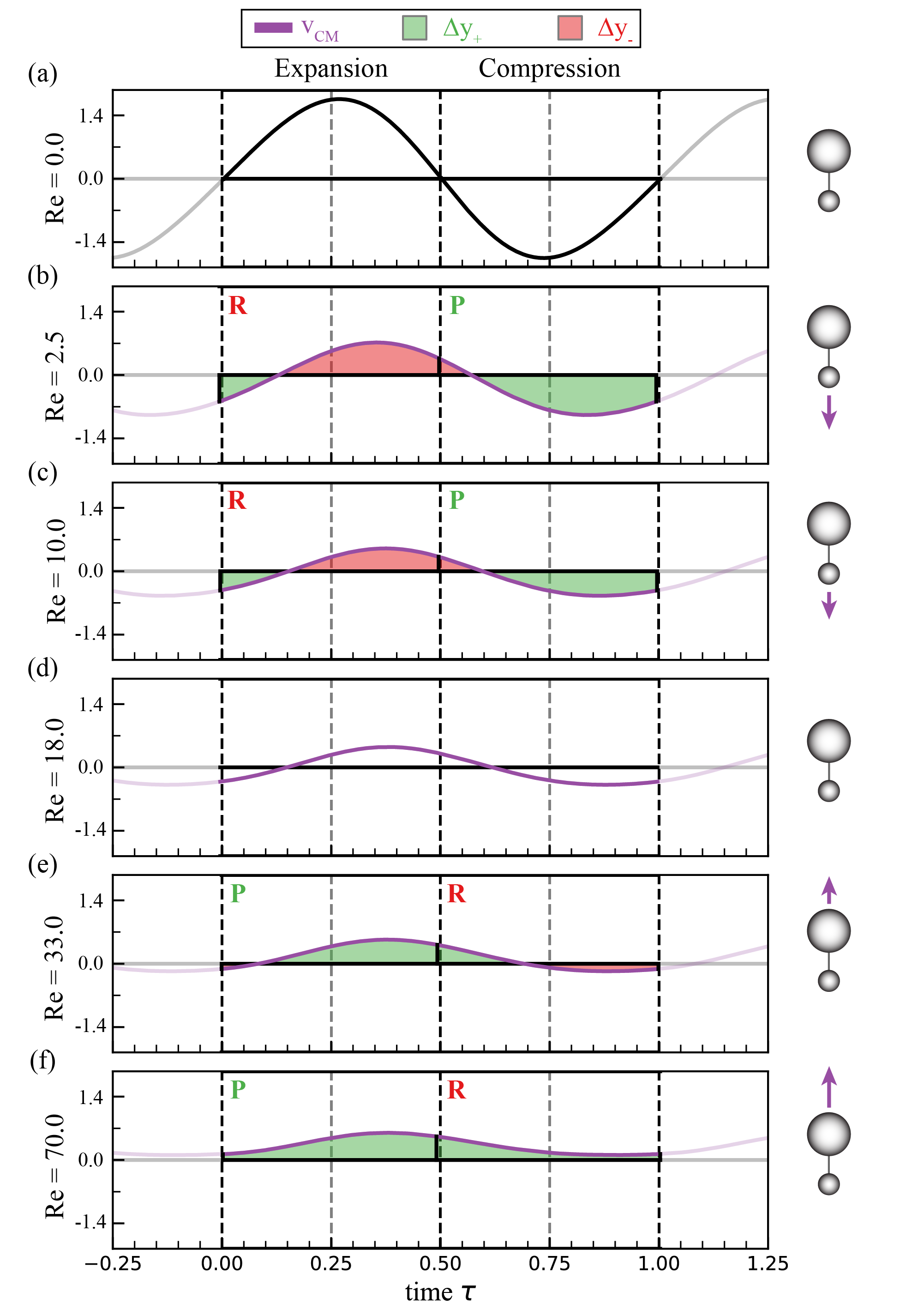}
\centering
\caption{The power and recovery stroke of the spherobot is determined by the movement of its appendage, the small sphere, represented by $v_{r}$. We define the power stroke to be when the small sphere moves opposite to the direction of net motion, $v_r \vavg < 0$. Vice versa, the recovery stroke is defined to be when the small sphere moves in the same direction as the net motion, $v_r \vavg > 0$. In this figure, the velocity of the spherobot, $v_{CM}$, is represented by the purple curves. The velocities are measured in (m/s). The mean swim direction $\vavg$ is indicated by the purple arrow in the accompanying spherobot schematics. The shaded areas under the $v_{CM}$ curve represent the displacements: in the mean swimming direction $\Delta y_{+}$ (green) and opposite to it $\Delta y_{-}$ (red). The power and recovery strokes for each swimmer are labeled by P and R, respectively. (a) The spherobot in Stokes flow. Here, there is zero net displacement. Therefore, there is no power or recovery stroke observed. (b) and (c) The spherobot swims net small-sphere-leading at Re=2.5. Its power stroke is during compression, and its recovery stroke is during expansion. (d) The spherobot does not swim and its net displacement is zero. Like Stokes flow, we do not observe a power or recovery stroke. (e) and (f) The spherobot swims large-sphere-leading at Re=33.0 and Re=70.0. Its power and recovery stroke are opposite of those observed for the small-sphere-leading spherobot. The power stroke occurs during expansion, and the recovery stroke occurs during compression.  
}
\label{PowRec}
\end{figure}

\subsection{Power and recovery}

To gain insight into the spherobot's motility mechanism in the two regimes, we divided its periodic motion into power and recovery strokes, a classical analysis for the motility of Stokesian swimmers~\cite{vogel1988life}. In living organisms, a common way to define power and recovery strokes is as follows. The power stroke occurs when the swimmer's appendage, i.e. the part of the swimmer that generates motion, moves opposite to the direction of the mean swim velocity $\vavg$, and the recovery stroke occurs when the appendage moves in the same direction as $\vavg$~\cite{Fish1996,Alben2013}. For example, one can imagine a human swimmer's breast stroke. The power stroke occurs when the swimmer's arms move back to propel the swimmer forward, and the recovery stroke occurs as the arms return to their original position. During the recovery stroke, the swimmer either moves backwards or slows down depending on the motility mechanism, Re, etc. It is also important to note that organisms with reciprocal strokes (stroke same forwards in time as backwards) cannot swim in Stokes flow meaning the power stroke is identical to the recovery stroke and the swimmer moves back and forth the same amount i.e. the scallop or the spherobot.

How does a power and recovery stroke emerge as Re increases from 0 to finite? And how do the notions of power and recovery strokes evolve as Re increases further? We aim to answer these questions for the spherobot. We view the large sphere as the body of the swimmer and the small sphere as its appendage. The justification of this is that the small sphere moves the most as it has a larger amplitude than the large sphere, see also~\cite{dombrowski2019transition}. Thus, we define the power stroke to be when the velocity of the small sphere and the average velocity of the CM over the whole cycle are in opposite directions $v_r\vavg<0$ and the recovery stroke when the velocity of the small sphere and the average velocity of the CM are in the same direction $v_r\vavg>0$. Note that $\vavg<0$ in the small-sphere-leading regime and $\vavg>0$ in the large-sphere-leading regime. 

In Fig.~\ref{PowRec}, we plot $v_{CM}$ (purple), the displacements in the same direction as $\vavg$ termed $\Delta y_{+}$ (green area) and opposite to it $\Delta y_{-}$ (red area), and indicate power (P) and recovery strokes (R) in each regime. For $\text{Re} = 0.0$, there was no distinction between power and recovery strokes because the spherobot does not swim $\vavg = 0.0$, Fig.~\ref{PowRec}(a). Connecting to the two regimes, in the small-sphere-leading regime, Fig.~\ref{PowRec}(b), the spherobot performs a power stroke during compression and a recovery stroke during expansion. The effect of inertia is already apparent: the curve has shifted in the time axis compared to Stokes flow, such that, early in the recovery stroke, the swimmer is still moving forward due to the power stroke. Similarly the swimmer is still moving backward early in the power stroke. As Re increases, the power and recovery strokes produce smaller displacements in both directions, see Fig.~\ref{PowRec}(c). As a result, the spherobot experiences less intense back-and-forth motion. Note that we do not see a further shift with respect to time. At the critical value where the transition in the swimming direction occurs ($\text{Re}=18.0$), expansion and compression generate smaller but equal displacements in both directions; so the spherobot remains stationary over a cycle, see Fig.~\ref{PowRec}(d). As Re increases further, the spherobot switches direction to swim large-sphere-leading, and now performs a power stroke during expansion and a recovery stroke during compression, Fig.~\ref{PowRec}(e,f). Its periodic motion is still prescribed and does not change, but the power and recovery strokes reverse. There is also a behavioral change in the recovery stroke. When $\text{Re} > 18$ but still close to the transition, the recovery stroke produces a backwards displacement (Fig.~\ref{PowRec}(e)), while for higher Re the recovery stroke does not produce a backward displacement and just slows down the swimmer (Fig.~\ref{PowRec}(f)). The power stroke, on the other hand, does not change much with Re and the maximum velocity remains approximately constant. This is a demonstration showing how the movement of a simple model swimmer is affected by the onset and gradual increase of inertia. 

\section{\label{sec:FluidFlow}Fluid Flows and Efficiency}

\begin{figure}
\includegraphics[width=0.75\columnwidth]{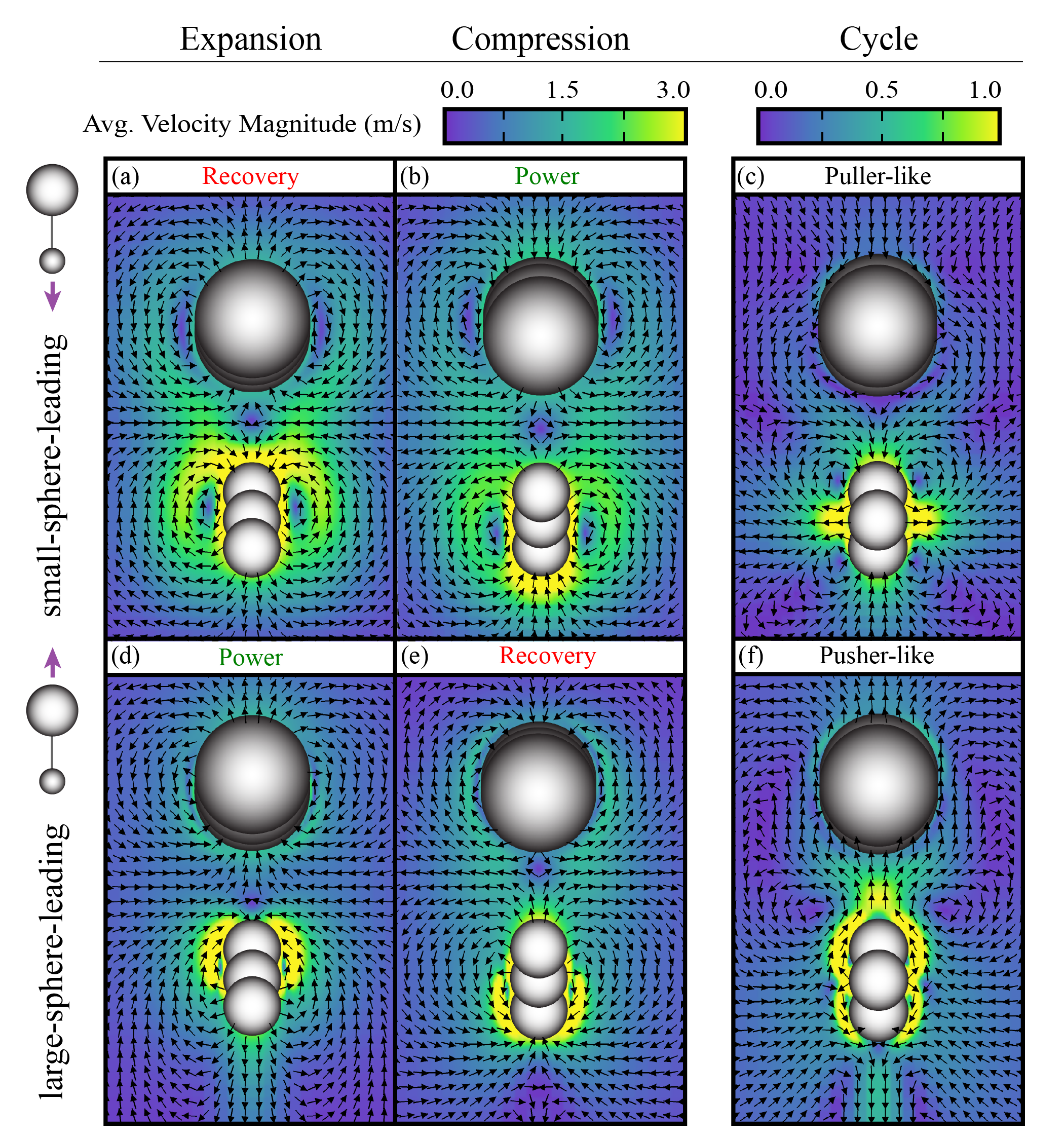}
\centering
\caption{The average velocity field of a small-sphere-leading (top row) and large-sphere-leading (bottom row) spherobot averaged over (left) expansion, (middle) compression, and (right) an oscillation cycle. Flow magnitudes are represented by the heat map, and the flow direction is indicated by the black arrows. (a) Flow field of small-sphere-leading spherobot averaged over expansion. The fluid flows outward along the swimming axis and inward perpendicular. (b) Flow field of small-sphere-leading spherobot averaged over compression. The flow is opposite to that of expansion, inward along the swimmer's axis and outward perpendicular. (c) Averaged over a whole cycle small-sphere-leading spherobot flow. The net flow is puller-like. (d) Flow field of large-sphere-leading spherobot averaged over expansion. The fluid flows outward along the swimming axis and inward perpendicular. (e) Flow field of large-sphere-leading spherobot averaged over compression. The flow is opposite to that of expansion, inward along the swimmer's axis and outward perpendicular. (f) Averaged over a whole cycle large-sphere-leading spherobot flow. The net flow is pusher-like.
}
\label{AvgFieldMB}
\end{figure}

To get more insight into how net motion was achieved we calculated the averaged fluid flow fields, and we present our findings for the two characteristic systems. Previously, we showed that in the small-sphere-leading regime the averaged flow over a cycle is puller-like i.e. the flow is pulled in toward the spheres along the swimming direction and is pushed out along the perpendicular, see Fig~\ref{AvgFieldMB}(c), while in the large-sphere-leading regime, the averaged flow over a cycle is pusher-like i.e. the flow is pushed out away from the spheres along the swimming direction and is pulled in along the perpendicular, see Fig~\ref{AvgFieldMB}(f)~\cite{dombrowski2019transition}. Here, we split the flows averaging over expansion and compression separately. The averaged flow during expansion resembles a pusher and during compression a puller for both swimming regimes. This makes intuitive sense as we expect the fluid to flow into the gap between the spheres during expansion and to be pushed out of the gap during compression. There is a competition between pusher and puller type flow, and depending on Re, either puller or pusher flow is more dominant, as evident by the difference in net flow fields (c) and (f). 

The presence of both puller and pusher flows during the cycle for both small-sphere leading (puller-like overall) and large-sphere-leading (pusher-like overall) regimes resemble the flows around living organisms in the following way. Stokesian swimmers such as \textit{Chlamydomonas} and sperm cells have been shown to oscillate between puller and pusher flows~\cite{Klindt2015} even though they are classified as a puller and pusher respectively based on the net far field flow.
Relating to power and recovery strokes in each regime, it is worth noting that the flow field that occurs during the power stroke is the one that dominates over the cycle, see Fig.~\ref{AvgFieldMB}. 

\begin{figure}
\includegraphics[width=0.5\columnwidth]{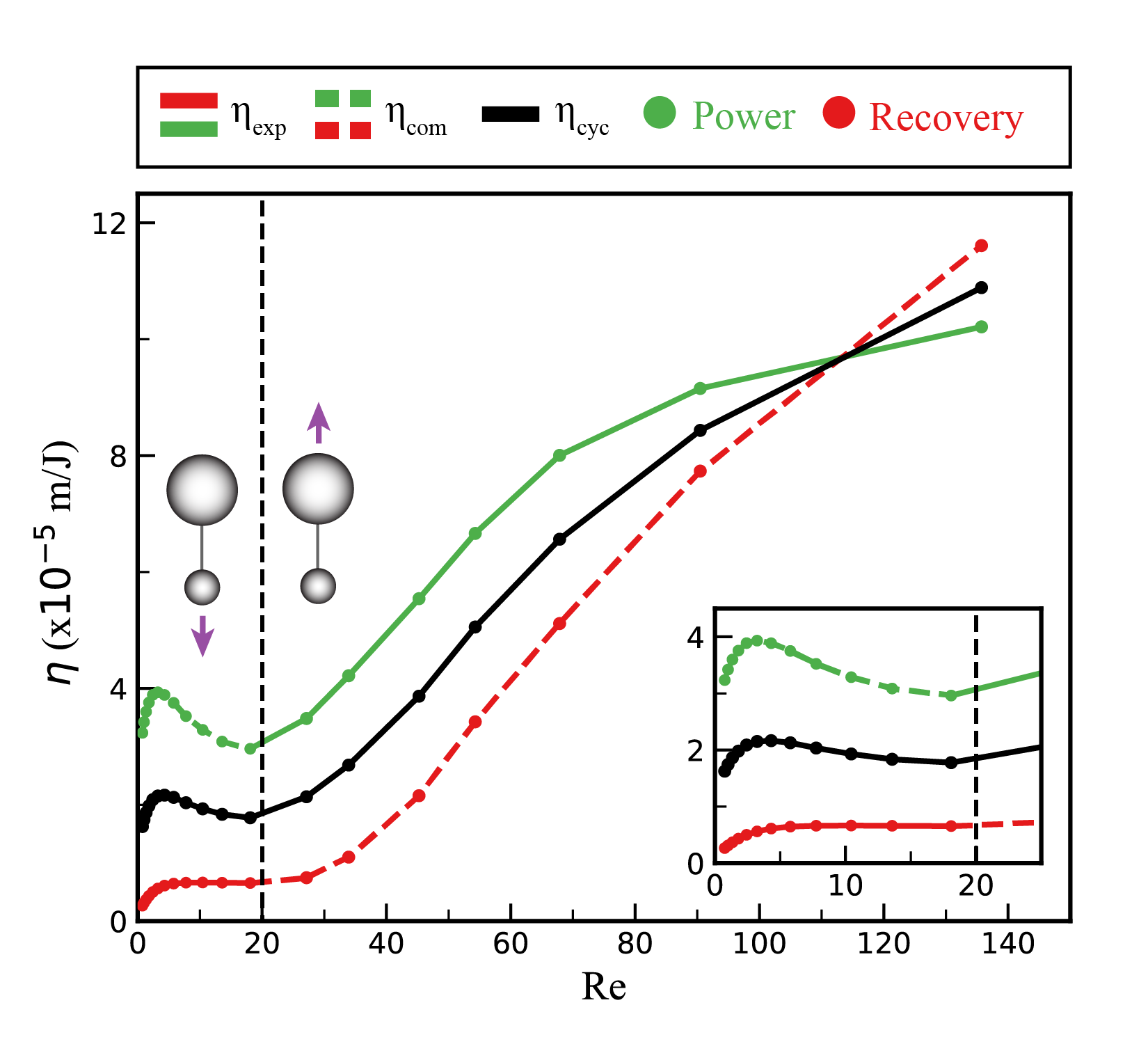}
\centering
\caption{Efficiency of a spherobot with parameters $A=0.18$m and $d_0=0.98$m as a function of Re where $\eta_{exp}$ (solid green and red), $\eta_{com}$ (dashed green and red), and $\eta_{cyc}$ (black) are depicted. Also shown are the efficiencies of the power (green) and recovery (red)  strokes. The inset shows a close up of the efficiencies in the small-sphere-leading regime. Here, the power stroke occurs during compression, the recovery stroke during expansion, and $\eta_{com} > \eta_{exp}$. Because there is a switch in swimming direction at $\text{Re} \approx 20$, the power and recovery strokes also switch. Now, $\eta_{exp} > \eta_{com}$. As expected, the power stroke is more efficient than the recovery, but up until $\text{Re} > 110$ for this configuration. There is also a trend in the spherobot's cycle efficiency where swimming large-sphere-leading is generally more efficient than swimming small-sphere-leading.
}
\label{Effic}
\end{figure}

We also calculated the efficiency of our swimmer in each regime during expansion $\eta_{exp}$, compression $\eta_{com}$, and the whole cycle $\eta_{cyc}$.
We defined the efficiency to be $\eta = \Delta y_+ / E$, where $\Delta y_+$ is the swimmer's total distance traveled in the net swimming direction and $E$ is the total energy added to the system. There was zero contribution to the efficiency when the spherobot moved opposite to its swim direction. In Fig.~\ref{Effic}, we calculated the efficiency of the spherobot with parameters $A=0.18$m and $d_0=0.98$m and plotted as a function of $\text{Re}$. In the small-sphere-leading regime, the spherobot was more efficient during compression (Fig.7 green dotted line) than expansion (Fig.7 red solid line), $\eta_{com} > \eta_{exp}$, i.e. it was more efficient to push fluid out from between the spheres than to pull it in. In the large-sphere-leading regime, expansion (Fig.7 green solid line) was more efficient than compression (Fig.7 red dotted line), $\eta_{exp} > \eta_{com}$ i.e. it was more efficient to pull fluid in between the spheres than to push it out. 
For most Re, the power stroke is more efficient than the recovery stroke.
However, at $\text{Re} \approx 110$ for this spherobot configuration, the recovery stroke becomes more efficient than the power stroke. In fact, the $\text{Re}$ where the recovery stroke becomes more efficient than the power stroke depends on the separation distance of the spheres, $d_0$. The larger the separation, the larger the Re where the recovery becomes the more efficient stroke, see Fig.~S11. We discuss possible explanations in the SI (section III). 

\begin{figure}
\includegraphics[width=0.75\columnwidth]{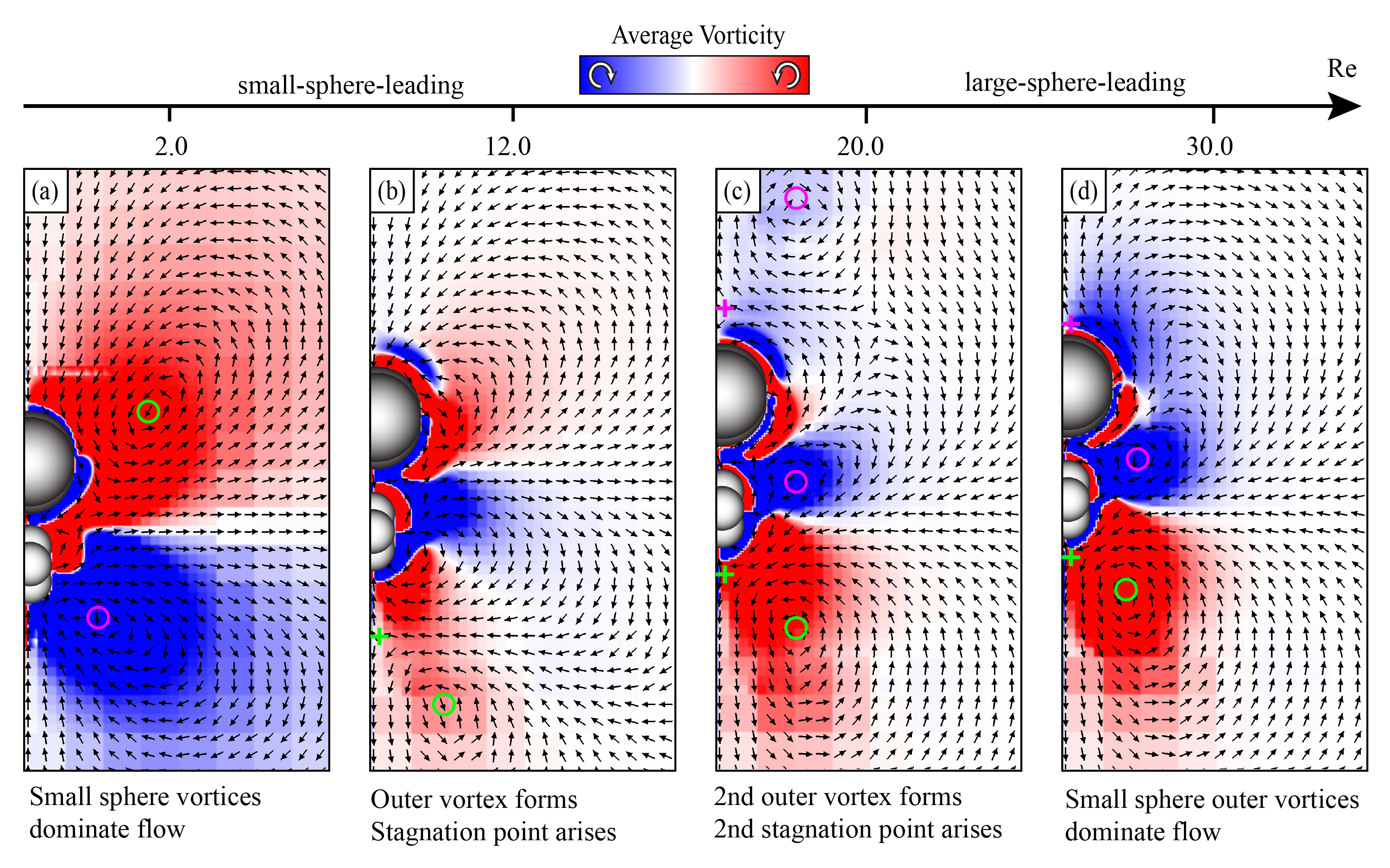}
\centering
\caption{
The fluid flow around the spherobot continuously evolves across Re. We provide four flows averaged over a cycle, from a spherobot with $d_0 = 0.75m$ and $A = 0.18m$, across both swimming regimes to highlight its evolution. Vortices of interest are identified with circles (O) and stagnation points are shown with a (+). Their colors are chosen to contrast background vorticity. (a) At $\text{Re} = 2.0$, the spherobot swims small-sphere-leading. We observe one pair of vortices from each sphere, and the small sphere's vortices dominate the surrounding flow. (b) At $\text{Re} = 12.0$, an outer vortex forms below the small sphere which rotates counter to the inner vortex. The flow direction change below the small sphere is shown to occur at the specified stagnation point. (c) At $\text{Re} = 20.0$, another outer vortex forms above the large sphere with an accompanying stagnation point. There is a competition between pushing and pulling the fluid both above and below the spheres, and the spherobot does not swim. (d) The spherobot now swims large-sphere-leading at $\text{Re} = 30.0$. The outer vortex above the large sphere disappears, and the flow merges with the outer vortex above the small sphere, pink circle. The outer vortex below the small sphere (green circle) remains and moves closer to the spherobot. The outer vortices generated from the small sphere movement dominate the surrounding fluid flow.
}
\label{ContFlow}
\end{figure}

We can accredit the motion of the spherobot to the continuous evolution in its averaged fluid flow over a cycle (steady streaming) across $\text{Re}$, see Fig.~\ref{ContFlow}. First, at low $\text{Re}$, Fig.~\ref{ContFlow}(a), the spherobot oscillations generate only one vortex layer. The flow pulls inward along the swimming axis and pushes outward along the perpendicular, see also Fig.6(c). Because of the asymmetry in the spherobot, there is a resulting asymmetry in fluid flow. The small sphere has a larger amplitude so its oscillation affects the surrounding flow farther away than the large sphere does. In fact, steady streaming flows theoretically scale as $A^2\omega/\nu$ \cite{riley2001}. Thus the averaged flow appears to be dominated by the small sphere so much that the large sphere acts almost as an obstacle. As a result, at the lower end of Re, the spherobot moves small-sphere-leading because the fluid below the small spheres pulls it more than fluid above. As Re increases, the inner vortex layer reduces in size and extent, and eventually an additional outer vortex layer forms only below the small sphere, see Fig.~\ref{ContFlow}(b). The outer vortex layer rotates counter to the inner vortex which creates a competition between pulling the spherobot down and pushing the spherobot up along its swimming axis (stagnation point). The spherobot slows down and approaches zero. An outer vortex layer above the large sphere develops at a higher $\text{Re}$ relative to the outer vortex below the small sphere i.e. Fig.~\ref{ContFlow}(c). When the spherobot is stationary at $\text{Re} = \text{Re}_c$, the inner vortex pulls the spherobot as strong as the outer vortex pushes it. As Re increases further, Fig.~\ref{ContFlow}(d), the outer vortex above the large sphere aligns its rotation with the outer vortex above the small sphere and it disappears. The small sphere's outer vortices become more and more dominant, and the spherobot is pushed up more by the fluid below the small sphere. The inner vortex becomes smaller, $\delta = \sqrt{\nu/\omega}$, and as a result the spherobot becomes more efficient in swimming large-sphere-leading. Thus, here is another example where we see how the spherobot's movement is due to a competition between pushing and pulling. 

\section{\label{sec:Disc}Discussion}

To summarize, we explored the spherobot's kinematics and its relationship with $\text{Re}$, amplitude $A$, and the equilibrium distance between spheres $d_0$ by collapsing the net displacements during expansion $\Delta y_{exp}$ and during compression $\Delta y_{com}$. In the small-sphere-leading regime, the spherobot performed a back-and-forth motion where it moved more in the direction of swimming during compression than in the opposite direction during expansion. The backwards motion disappeared as Re increased and the spherobot moved in the direction of swimming during expansion and slowed down during compression. We categorized the spherobot's swimming into power and recovery strokes. The swim stroke itself did not change, however, due to the change in swim direction, the power and recovery strokes switched. We looked at the individual sphere's velocities and identified regions of slip where both spheres and the spherobot's CM moved in the same direction. We noticed the slip direction at the end of the power stroke was always in the same direction as the net swimming. We analyzed the flow fields for a small-sphere-leading and large-sphere-leading spherobot. Much like living organisms, there was a competition between puller and pusher type flow throughout the cycle. When averaged over the whole cycle, the flow that dominated was the one that occurs during the power stroke. We calculated the efficiencies of the spherobot over the cycle as well as during expansion and compression separately. We determined that in the small-sphere-leading regime, it was more efficient to push fluid out of the gap between the spheres than to pull fluid inward; the opposite was true for most Re in the large-sphere-leading regime. There was additional complexity in the efficiency in the large-sphere-leading regime where we found that at high Re depending on $d_0$ the recovery stroke was more efficient than the power stroke.

We stress the importance in understanding motility and its complexity at intermediate Re. Recent studies have reported other model swimmers which can switch their swim direction based on internal or external stimuli, see e.g. passively flapping plate~\cite{zhang2010locomotion} and asymmetric dumbbell shaker~\cite{Collis2017} respectively. It remains open to see whether other model swimmers at intermediate Re show similar behavior, and what kind of classifications can be made. 

Another area of importance is how collective behavior emerges from the nonlinearities that arise when many mesoscale organisms swim together. Are there systems where, say, two organisms individually swim in one preferred direction, but together as a collective swim differently? Finally, from an applications standpoint, it is important to understand the underlying physical mechanisms behind motility at intermediate Re, impacting the design of artificial swimmers, drones, and inertial microfluidics.

\textbf{Acknowledgments}
D.K. and T.D acknowledge the National Science Foundation, grant award DMR-1753148.

\bibliography{references}

\end{document}